\documentclass[aps,preprint,superscriptaddress]{revtex4}
\usepackage{graphicx}
\begin{document}

\title{Effect of strain on stripe phases in the Quantum Hall regime}

\author{Sunanda~P. Koduvayur}
\author{Yuli Lyanda-Geller}
\author{Sergei Khlebnikov}
\author{Gabor Csathy}
\affiliation{Department of Physics and Birck Nanotechnology Center, Purdue
University, West Lafayette, IN 47907 USA}
\author{Michael~J. Manfra}
\affiliation{Department of Physics and Birck Nanotechnology Center, Purdue
University, West Lafayette, IN 47907 USA}
\affiliation{School of Materials Engineering and School of Electrical and Computer Engineering, Purdue University,
West Lafayette, IN 47907 USA}

\author{Loren~N. Pfeiffer}
\author{Kenneth~W. West}
\affiliation{Department of Electrical Engineering, Princeton University,
Princeton, NJ 08544 USA}

\author{Leonid~P. Rokhinson}
\affiliation{Department of Physics and Birck Nanotechnology Center, Purdue
University, West Lafayette, IN 47907 USA}

\begin{abstract}
Spontaneous breaking of rotational symmetry and preferential orientation of stripe phases in the quantum Hall regime has attracted considerable experimental and theoretical effort over the last decade. We demonstrate experimentally and theoretically that the direction of high and low resistance of the two-dimensional (2D) hole gas in the quantum Hall regime can be controlled by an external strain. Depending on the sign of the in-plane shear strain, the Hartree-Fock energy of holes or electrons is minimized when the charge density wave (CDW) is oriented along $[110]$ or $[1\bar{1}0]$ directions. We suggest that shear strains due to internal electric fields in the growth direction are responsible for the observed orientation of CDW in pristine electron and hole samples.
\end{abstract}

\date{May 18, 2010}

\maketitle

Interplay between kinetic energy and electron-electron interactions in two-dimensional electron gases in magnetic fields leads to a rich variety of possible ground states, ranging from incompressible Laughlin liquids, the Wigner crystal, charge density waves (CDW) to exotic non-Abelian anyonic states. The possibility of the formation of a CDW state had been suggested\cite{fukuyama79} even before the discovery of the quantum Hall effect, and later it was predicted that a CDW should be the ground state for partially occupied high Landau levels\cite{fogler96,koulakov96}. Experimentally, anisotropic magnetoresistance (AMR) and re-entrant QH phases have been observed in 2D electron\cite{lilly99,du99} and hole\cite{shayegan00,manfra07} gases. The majority of experiments have been conducted on samples grown on (001) GaAs. Unexpectedly, the CDW was found to be consistently oriented along [110] crystallographic direction in these samples, a surprising fact considering isotropic nature of the wave functions on the high symmetry (001) surface.

Search for the physical origin of the broken symmetry and the observed preferential orientation of stripe phases has been actively pursued experimentally and theoretically over the past decade. Reduced symmetry of the interface was suggested \cite{kroemer99} as a factor which introduces an anisotropy of the effective mass\cite{volkov00} or of the cyclotron motion\cite{rosenow01}. However, single-particle effects associated with these anisotropies seem unlikely to be responsible for the large magnitude and strong temperature dependence of the resistance\cite{fogler-review}. Later work showed\cite{cooper01} that the precise symmetry of the 2D gas confining potential is also unimportant, and micron-scale surface roughness does not correlate with the stripe orientation. There have been theoretical suggestions\cite{rashba87,sherman95,fil00} that anisotropic correction to electron-electron interactions arising from elastic and piezoelectric effects can be responsible for the resistance anisotropy. While the free energy of the CDW is minimized in the vicinity of [110] and $[1\bar{1}0]$ directions, those theories cannot explain why these directions are inequivalent. Progress has been made in understanding the effect of the in-plane magnetic field, which has been shown to influence orientation of stripes\cite{pan99,lilly99a}, the effect being explained by the field-induced anisotropy of the exchange potential\cite{jungwirth99a,stanescu00}. However, naturally existing preference for $[110]$ orientation of the CDW in purely perpendicular field, the same for electron and hole samples, remained unresolved.

In this work we show both experimentally and theoretically that strain breaks the cylindrical symmetry of electron-electron interactions in magnetic field and results in a preferred orientation of the CDW. We show experimentally that externally applied shear strain can enhance or reduce anisotropy of the resistance and switch low and high resistance axes. Our theory shows that depending on the sign of the in-plane shear strain, the Hartree-Fock energy is minimized when the CDW is oriented along $[110]$ or $[1\bar{1}0]$ directions. We suggest that shear strains due to internal electric fields in the growth direction caused by mid gap Fermi level pinning at the sample surface are responsible for the observed preferred orientation of the CDW in pristine electron and hole samples. Finally, by applying uniaxial strain we are able to induces a stripe phase even at a filling factor $\nu=5/2$, with a CDW winning over other QH states.

\begin{figure}[t]
\def\ffile{stripes}
\includegraphics[scale=0.32]{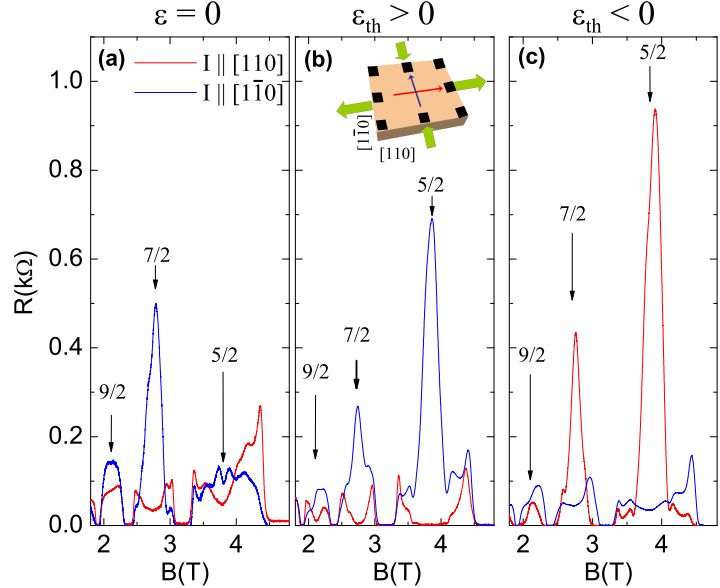}
\caption{Magnetoresistance is plotted as a function of $\bf B$ for the current aligned with $[110]$ (red) and $[1\bar{1}0]$ (blue) directions in strained and unstrained samples. In a) no external strain is applied, (b) thermally-induced tensile strain is along $[110]$, and  along $[1\bar 10]$, $\varepsilon_p=0$. Inset shows sample schematic, red and blue arrows show current, and green arrows show strain.}
\label{\ffile}
\end{figure}

Samples were fabricated in the van der Pauw geometry from carbon doped GaAs quantum well heterostructure grown on $(001)$ GaAs \cite{manfra05,manfra07}. From the low field Shubnikov de-Haas oscillations, the hole density is $2.25\times 10^{11}$ cm$^{-2}$, and the mobility $0.8\times 10^{6}$ cm$^{2}$/Vs is determined at the base temperature 10mK. Some samples were thinned to $150\ \mu$m and glued on a multilayer PZT (lead zirconate titanate) ceramic actuator with samples' [110] or $[1\bar 10]$ crystallographic axis aligned with the polarization axis of the PZT. Application of voltage $V_p$ to the actuator induces in-plane shear strain in the sample $\varepsilon_{p}/V_p=2.8\times 10^{-7}$ V$^{-1}$ and small uniform bi-axial strain. The total shear strain $\varepsilon=\varepsilon_{th}+\varepsilon_{p}$ also includes a residual strain $\varepsilon_{th}$ due to anisotropic thermal coefficient of the actuator, which depends on the $V_p$ during cooldown. To insure that voltage on the actuator does not induce charge modulation in the attached sample we insert a thin metal foil between the sample and the PZT. The foil was also used as a back gate to adjust 2D gas density which has a weak dependence on strain due to difference in piezoelectric coefficients of GaAs and AlGaAs\cite{fung97} (density changes by 3\% for the maximum voltage span on the PZT).

\begin{figure}[t]
\includegraphics[scale=0.50]{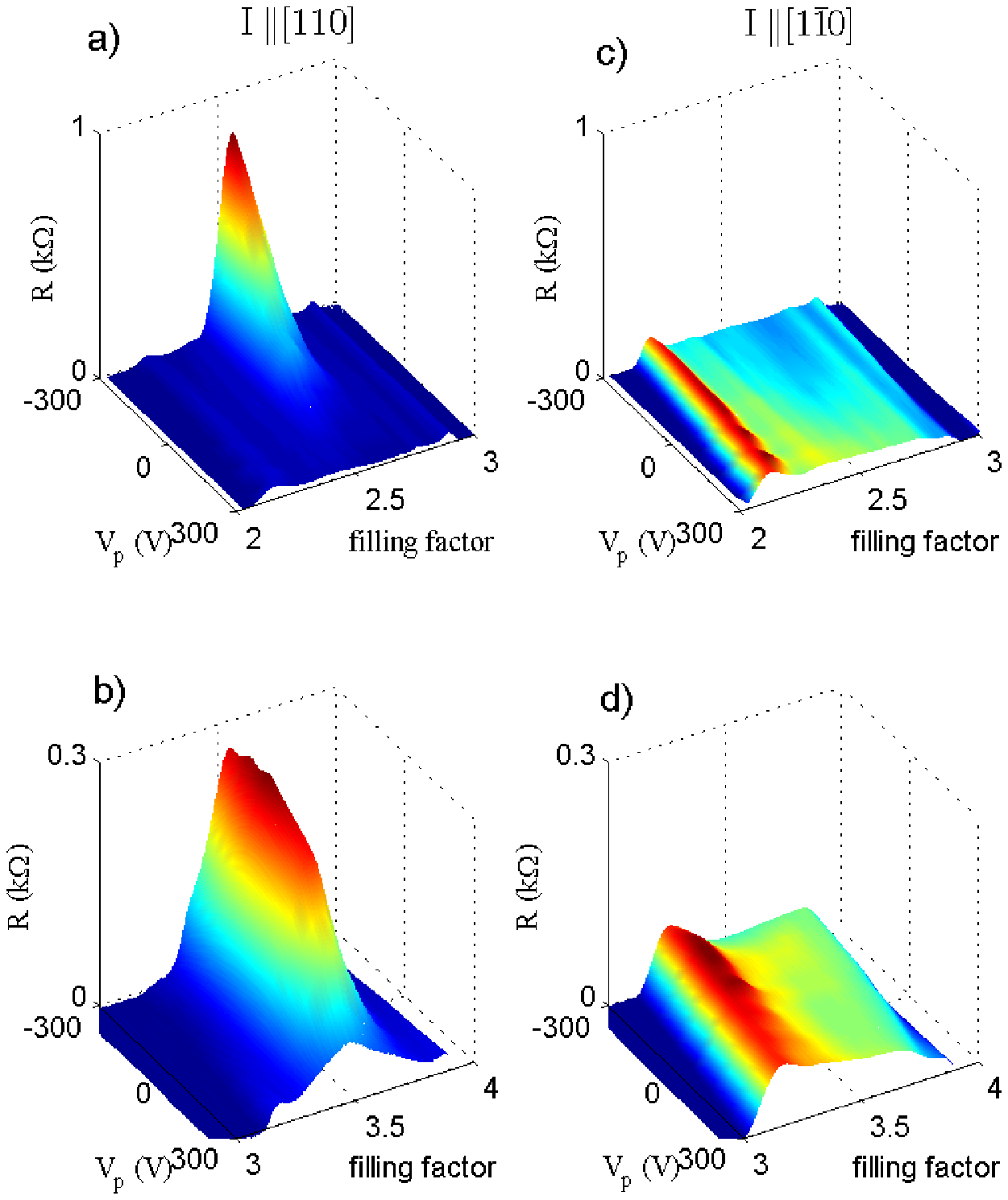}
\includegraphics[scale=0.30]{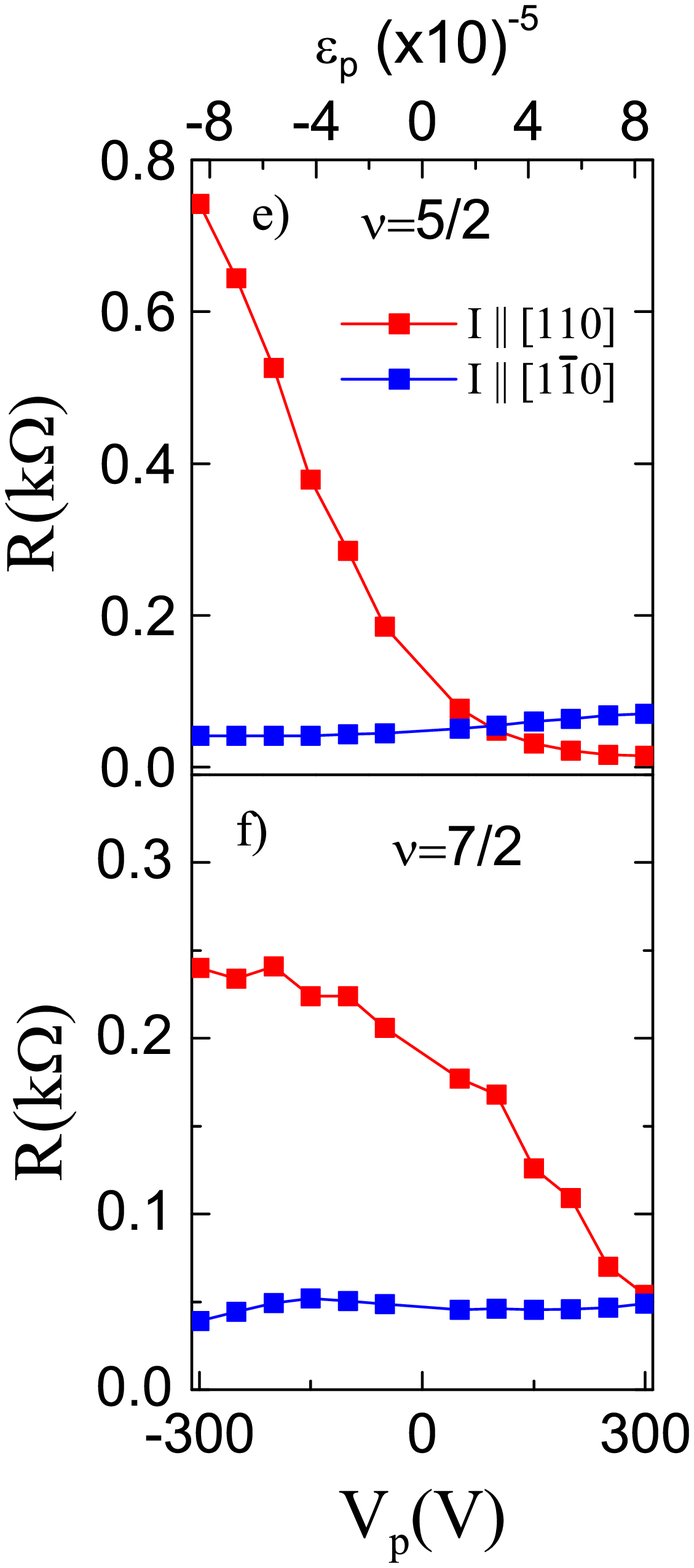}
\caption{Strain dependence of the anisotropic magnetoresistance. (a-d)Magnetoresistance in the vicinity of $\nu=7/2$ and 5/2 for $I\|[110]$ and $I\|[1\bar10]$ as a function of voltage on the piezoelectric actuator $V_p$. In (e-f) magnetoresistance at $\nu=7/2$ and 5/2 is extracted. On the top axis, $V_p$ is converted to the piezo-induced shear strain in the heterostructure; actual strain includes thermally-induced offset.}
\label{strain-3d}
\end{figure}

Magnetoresistance for pristine (i.e., not attached to an actuator) sample is shown in Fig.~\ref{stripes}a. States at $\nu=7/2$ and $11/2$ are highly anisotropic with low resistance direction along [110], while states at $\nu=5/2$, 9/2 and $13/2$ are almost isotropic, consistent with the previous study\cite{manfra07}. In Fig.~\ref{stripes}b,c similar traces are shown for large $\varepsilon=\varepsilon_{th}\gtrsim+10^{-4}$ and $\varepsilon=\varepsilon_{th}\lesssim-10^{-4}$ ($\varepsilon_p=0$). For $\varepsilon_{th}>0$ the anisotropy is enhanced compared to the unstrained sample, with states at $\nu=5/2, 9/2$ and $13/2$ becoming anisotropic and resistance for $I \| [110]$ approaching zero for half-filled Landau levels. For $\varepsilon_{th}<0$ the low and high resistance axes are switched. Here strain also leads to strong anisotropy at $\nu=5/2$ with high resistance axis along [110] direction.

Residual strains $\varepsilon_{th}$ in Fig.~\ref{stripes}b,c are larger than the {\it in situ} adjustable strain $\varepsilon_{p}$. In Fig.~\ref{strain-3d} we analyze AMR for a sample cooled with $V_p=-150$ V, aiming for $\varepsilon_{th}\sim0$. Magnetoresistance as a function of $\varepsilon_{p}$ is plotted near $\nu=5/2$ and 7/2. At $V_p<0$ magnetoresistance is highly anisotropic. For the high resistance direction, the resistance $R_{I\|[110]}$ strongly depends on $\varepsilon_p$ and decreases by a factor of 50 (4.4) at $\nu=5/2$ (7/2) as $V_p$ is varied from -300 V to 300 V. $R_{I\|[1\bar10]}$ increases only 1.7 (1.3) times. At $V_p>100$ V, the resistance at $\nu=5/2$ is isotropic, with no maxima for either current direction, as in unstrained sample. Thus, in the range of small strains,  the CDW is not a ground state at $\nu=5/2$, consistent with observations in unstrained samples. From the data we conclude that $\varepsilon<0$ within the adjustable range of $V_p$, because $R_{I\|[110]}>R_{I\|[1\bar10]}$ at $\nu=7/2$. Continuous evolution of $R_{I\|[110]}$ and $R_{I\|[1\bar10]}$  is consistent with continuous change reported in in-plane magnetic field \cite{lilly99a}.

Having presented experimental results of strain on the resistance in the QH regime, we now develop the Hartree-Fock theory of CDW by extending theoretical work \cite{aleiner95} to anisotropic 2D systems. Strain results in sizable modification of the 2D single-particle spectra, and of the many-body wavefunctions. For either electrons and holes, the 2D Hamiltonian in magnetic field ${\mathbf H}= \nabla\times{\mathbf A}$ is
\begin{equation}
{\cal H}_{2D}= \frac{(p_x-\frac{e}{c}A_x)^2} {2m_x}+ \frac{(p_y-\frac{e}{c}A_x)^2}{2m_y},
\label{spectrum}
\end{equation}
where $x\| [110]$, $y \| [1\bar{1}0]$ are the principal axes of the reciprocal mass tensor, with $m^{-1}=(m_x^{-1}+m_y^{-1})/2$ and $\mu^{-1}=(m_x^{-1}-m_y^{-1})/2$ being isotropic and anisotropic parts. For holes in a III-V material quantized along the $(001)$ direction $m^{-1}= - (\gamma_1+\gamma_2+ \alpha\gamma_3)/m_0$, where $\gamma_1,\gamma_2,\gamma_3$ are negative constants defining the bulk hole spectra \cite{birpikus74,luttinger56} and the numerical coefficient $\alpha$ is defined by these constants\cite{nedorezov71}. For the anisotropic part induced by the shear strain $\varepsilon=\varepsilon_{xx}=-\varepsilon_{yy}$ our result is $\mu^{-1} =\gamma_3 \sqrt{3}d\varepsilon/[\gamma_2(\pi\hbar/a)^2]$, where $d$ is the deformation potential \cite{birpikus74}, and $a$ is the quantum well width (we assume an infinite rectangular QW). For electrons, $m$ is the 3D effective electron mass $m_c$, while for the anisotropic part our result, obtained in the third-order perturbation theory, is $\hbar^2/2\mu= -P^2d\varepsilon/\sqrt{3}E_g^2$, where $P$ is the Kane band coupling parameter\cite{kane57} and $E_g$ is the band gap. Note that the sign of the strain-induced term for electrons is opposite compared to that for holes.

To find the single particle wavefunctions we define new coordinates, $x^{\prime}= x\sqrt{m_x/m'}$, $y^{\prime}= y\sqrt{m_y/m'}$, where $m'=\sqrt{m_xm_y}$. In these coordinates, ${\cal H}_{2D}=\frac{1}{2m'} (\mathbf{p}^{\prime}- \frac{e}{c}\mathbf{A}^{\prime})^2$ is isotropic, and the wavefunctions are the usual LLs wavefunctions with degeneracy in the guiding center coordinate $X^{\prime}=k_y^{\prime}l^2$, where $l=(\hbar c/eH)^{1/2}$ is the magnetic length. According to the physical picture of Aleiner and Glazman \cite{aleiner95}, the electron-electron interactions in a system with partially filled high topmost LL can be considered as interactions via the Coulomb potential with an effective dielectric constant defined by the electrons of the fully filled LLs. In the deformed coordinates, for $N\gg r_s^{-1}\gg 1$, where $N$ is the LL index, $r_s=(\pi na_B^2)^{-1}$, $n$ is the 2D carrier density, and $a_B=\hbar^2/me^2$ is the Bohr radius, the low energy physics of the 2D electron liquid in weak magnetic field is thus described by the effective Hamiltonian ${\cal H}_{eff} = \frac{1}{L_xL_y}\sum_{\mathbf{q}^{\prime}} \rho(\mathbf{q}^{\prime}) v(\mathbf{q}^{\prime}) \rho(-\mathbf{q}^{\prime})$, where $L_x,L_y$ give the size of the sample,
\begin{equation}
v(\mathbf{q}^{\prime})= \frac{2\pi e^2}{\kappa_0\kappa(\mathbf{q}^{\prime}) \sqrt{(q_x^{\prime 2}) \frac{m_x}{m}+(q_y^{\prime 2})\frac{m_y}{m}}}
\end{equation}
is the Fourier component of the renormalized electron-electron interaction potential, $\kappa_0$ is the background dielectric constant, and $\kappa(\mathbf{q}^{\prime})$ is the effective dielectric constant. An expression for $\kappa$ valid at $1/(\sqrt{2N+1}l)< Q^{\prime}< k_F$, where $k_F$ is the Fermi wavevector, is  $\kappa(\mathbf{q}^{\prime})=1+2/qa_B$\cite{kukushkin88}. Note that this $\kappa(\mathbf{q}^{\prime})$ is isotropic in the original coordinates but anisotropic in the deformed ones. Finally, $\rho(\mathbf{q}^{\prime})$ is the Fourier-component of charge density of the partially filled LL $N$, $ \rho(\mathbf{q}^{\prime})= \sum_{X^{\prime}} \alpha_N(\mathbf{q}^{\prime}) e^{-iq_x^{\prime} (X^{\prime}-q_y^{\prime}l^2/2)} a_{X^{\prime}}^{\dagger}a_{X^{\prime}-q_y^{\prime}l^2},$ where $\alpha_N(\mathbf{q}^{\prime})=L_N^0(\frac{\mathbf{q}^{\prime 2} l^2}{2}) \exp{(-\frac{\mathbf{q}^{\prime 2} l^2}{4})}$, $L_N^0(x)$ is the Laguerre polynomial, and $a_{X^{\prime}}^{\dagger}$ ($a_{X^{\prime}}$) is the creation (annihilation) operator for a hole with guiding center at $X'$ in the topmost LL. Defining the CDW order parameter $\Delta ({\mathbf q}')=\frac{2\pi l^2}{L_xL_y}\sum_{X'} a_{X'+q_y'l^2/2}^{\dagger} a_{X'-q_y'l^2/2}$, we obtain the Hartree-Fock energy as
\begin{equation}
E_{HF}= \frac{1}{2\pi l^2\nu_N}\sum_{\mathbf{q}^{\prime}\ne 0} [u_H(\mathbf{q}^{\prime})-u_{ex}(\mathbf{q}^{\prime})] \Delta(\mathbf{q}^{\prime})
\Delta(-\mathbf{q}^{\prime}),
\end{equation}
where $0\leq \nu_N\leq 1$ is the filling of the topmost LL. The Fourier transforms of the Hartree and exchange potentials are, respectively, $u_H(\mathbf{q}^{\prime})=v(\mathbf{q}^{\prime}) [\alpha_N(\mathbf{q}^{\prime})]^2$ and
\begin{equation}
u_{ex}(\mathbf{q}^{\prime})= \frac{2\pi l^2}{L_xL_y} \sum_{\mathbf{Q}^{\prime}} u_H(\mathbf{Q}^{\prime})
\exp{[il^2(-Q_x^{\prime}q_y^{\prime}+Q_y^{\prime}q_x^{\prime})]}.
\label{uex}
\end{equation}
Note that the two potentials are related by a Fourier transform but the arguments in the Fourier transform of the Hartree potential are transposed relative to the arguments of $u_{ex}$. While this transposition is not important in the isotropic case \cite{fogler96}, taking it into account here is crucial for finding the preferred orientation of the CDW modulation.

\begin{figure}
\def\ffile{ratio}
\includegraphics[scale=0.35]{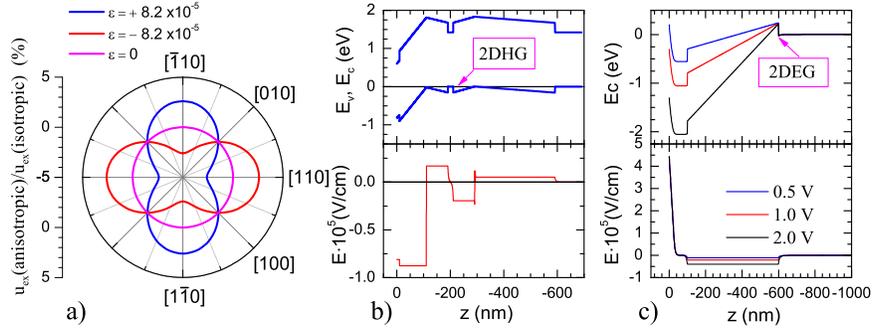}
\caption{a) The ratio of isotropic and anisotropic parts of the hole exchange potential (Eq.~\ref{uex}) for three values of shear strain $\varepsilon$. b) Self-consistent calculations of the band profile and internal electric fields in the studied wafer. c) Modeling of a HIGFET structure from \cite{zhu02}.}
\label{\ffile}
\end{figure}

Applying an analysis similar to that in \cite{fogler96}, we see that when the $L_N^0(x)$ in $\alpha_N(\mathbf{q}^{\prime})$ is zero, $u_H(\mathbf{q}^{\prime})$ is also zero, so that $E_{HF}<0$, as given by the exchange contribution. The system becomes unstable with respect to the formation of the CDW with a wavevector close to $\mathbf{q}^{\prime}$. However, in contrast to the isotropic system, in which the direction of the CDW is chosen spontaneously, in the presence of strain the smallest $\mathbf{q}^\prime$, which corresponds to the direction with the largest mass, gives the largest value of the exchange and the lowest HF energy. In Fig.~\ref{ratio}a, we plot the ratio of the anisotropic and isotropic parts of the hole exchange potential at $d=-5.4eV$, $\gamma_1=-6.8$, $\gamma_2=-2.1$, $\gamma_3=-2.9$, and $\alpha=0.4$. Note that here the anisotropy reaches 3\% for strains of $10^{-4}$. The CDW near the half filling of the $N$-th LL naturally results in stripes with alternating $\nu_N=0$ and $\nu_N=1$, which, in turn, translates into low resistance direction along the stripes and high resistance direction perpendicular to the stripes. Thus, in the presence of $\varepsilon\neq0$, the CDW has a preferential direction defined by the sign of the strain, consistent with the experimental results. While the theory is valid, strictly speaking, in high Landau levels it appears to describe experiments even at $N=1$, the lowest LL for which $L_N^0(x)$ has a zero.

The preceding analysis suggests that internal strain may be responsible for the observed orientation of stripes in pristine samples, where no external strain is applied.  GaAs is a piezoelectric material and any electric field in the $\hat z$ direction results in an in-plane shear strain $\varepsilon=d_{14}E_z$, where $d_{14}=-2.7\cdot10^{-10}$ cm/V. A calculated band diagram for our samples is shown in Fig.~\ref{ratio}b. Inside the QW $E_z<2\cdot10^4$ V/cm and results in strain too small to orient the stripes. $E_z$ on both sides of the QW, caused by doping, is also small and, in our samples, odd in $z$. However, in all GaAs samples there is a large field near the surface of the wafer due to the pinning of the Fermi energy near mid-gap. This surface charge-induced field is $\sim-10^6$ V/cm and the corresponding strain is $\varepsilon\sim3\cdot10^{-5}$. If transmitted to the QW region, this strain has the correct sign and magnitude to explain the observed orientation of stripes in pristine samples. To show that transmission of strain does indeed occur, we consider the minimal model in which this effect is present. The free energy of the model is given by ${\cal F} = {\cal F}_{\rm el} + 2 \beta E_z \varepsilon$, where the elastic free energy is ${\cal F}_{\rm el} = \frac{1}{2} \lambda \left[ (\partial_x u_y + \partial_y u_x)^2  + (\partial_z u_x)^2 +  (\partial_z u_y)^2 \right]$, $u_x$ and $u_y$ denote displacements, $\varepsilon = \frac{1}{2} (\partial_x u_y + \partial_y u_x)$ is the shear strain,  $x,y,z$ now correspond to the $[100]$, $[010]$, $[001]$ directions, $\beta$ is the piezoelectric constant, and $\lambda$ is the elastic one. In realistic devices, $E_z(x,y,z)$ is nonuniform in $(x,y)$ plane due to charge fluctuations on the surface and in the doping layer. For illustration, we consider a cylindrical sample of radius $L$ and height $d$ with cylindrically symmetric $E_{z}(r,z)=\sum_n E_n(z) J_0(q_n r)$, where $r=(x^2+y^2)^{1/2}$, $J_0$ is the Bessel function, and $q_n$ are quantized by the condition $J_0'(q_n L)=0$. The solution for the strain is
\begin{equation}
\label{strain}
\varepsilon(r,z) = \frac{\beta}{2\lambda} \sum_{n} \frac{q_n^2}{\partial_z^2 - q^2_n} E_n(z) J_0(q_n r) + \varepsilon_b(r,z),
\end{equation}
where $\varepsilon_b$ is localized near $r =L$. If $E_z(r,z)$ varies with $r$ at some characteristic scale $R$, a typical $q_n$ is of order $1/R$. Then, according to Eq.~\ref{strain}, the corresponding component of the strain propagates largely undiminished over distances  $z \sim R$ from the region where $E_n(z)$ is large (sample surface). As a result, macroscopic regions with sizable strain exist throughout the QW region.

The presence of internal strain also explains orientation of stripes found in 2D electron gases. A typical band diagram of an electron sample is similar to that shown in Fig.~\ref{ratio}b, but is inverted relative to the Fermi level with both the surface electric field and the shear strain changing sign. However, the anisotropic term in Eq.~\ref{spectrum} also has opposite sign for electrons and holes, so that the sign of the anisotropic term in electron and hole exchange is the same. Thus, for both holes and electrons, a surface field will orient the CDW along $[110]$, as seen in experiments. With our model, we can also explain the reorientation of stripes as a function of density observed in a HIGFET (Heterojunction-Insulated Gate Field Effect Transistor) \cite{zhu02}. At low gate voltages (low densities), shear strain will be dominated by the surface field, see Fig.~\ref{ratio}c. At large gate voltages (2 V corresponds to  $3\cdot10^{11}$ cm$^{-2}$), the electric field across the AlGaAs barrier becomes large enough to change the sign of the strain in the 2D gas region, thus reorienting the stripes.

In summary, we have shown experimentally that resistance anisotropy in the QH regime can be controlled by external strain. Theoretically, we have traced this effect to a strain-induced anisotropy of the exchange interaction and a competition between the internal and external strain, which defines the preferred direction of CDW. In general, any factor that brings in a crystallographic anisotropy of the effective mass gives rise to a crystallographic anisotropy of the Hartree-Fock energy of the CDW state (this is the case, for example, in hole gases grown on low-symmetry (311) GaAs \cite{shayegan00}). Our analysis suggests that, for heterostructures grown in the high-symmetry (001) surfaces, piezoelectricity due to surface electric fields is the largest source of such anisotropies. We underscore that, although the anisotropy of electron Hartree-Fock energy is two orders of magnitude smaller than that for holes, it must still choose a preferential direction for the CDW of guiding centers. Therefore, the preferential direction of the resistance anisotropy in pristine (001) samples appears to be universally dictated by internal strain.

\bibliography{rohi}

\begin{thebibliography}{28}
\expandafter\ifx\csname natexlab\endcsname\relax\def\natexlab#1{#1}\fi
\expandafter\ifx\csname bibnamefont\endcsname\relax
  \def\bibnamefont#1{#1}\fi
\expandafter\ifx\csname bibfnamefont\endcsname\relax
  \def\bibfnamefont#1{#1}\fi
\expandafter\ifx\csname citenamefont\endcsname\relax
  \def\citenamefont#1{#1}\fi
\expandafter\ifx\csname url\endcsname\relax
  \def\url#1{\texttt{#1}}\fi
\expandafter\ifx\csname urlprefix\endcsname\relax\def\urlprefix{URL }\fi
\providecommand{\bibinfo}[2]{#2}
\providecommand{\eprint}[2][]{\url{#2}}

\bibitem[{\citenamefont{Fukuyama et~al.}(1979)\citenamefont{Fukuyama, Platzman,
  and Anderson}}]{fukuyama79}
\bibinfo{author}{\bibfnamefont{H.}~\bibnamefont{Fukuyama}},
  \bibinfo{author}{\bibfnamefont{P.~M.} \bibnamefont{Platzman}},
  \bibnamefont{and} \bibinfo{author}{\bibfnamefont{P.~W.}
  \bibnamefont{Anderson}}, \bibinfo{journal}{Phys. Rev. B}
  \textbf{\bibinfo{volume}{19}}, \bibinfo{pages}{5211} (\bibinfo{year}{1979}).

\bibitem[{\citenamefont{Fogler et~al.}(1996)\citenamefont{Fogler, Koulakov, and
  Shklovskii}}]{fogler96}
\bibinfo{author}{\bibfnamefont{M.~M.} \bibnamefont{Fogler}},
  \bibinfo{author}{\bibfnamefont{A.~A.} \bibnamefont{Koulakov}},
  \bibnamefont{and} \bibinfo{author}{\bibfnamefont{B.~I.}
  \bibnamefont{Shklovskii}}, \bibinfo{journal}{Phys. Rev. B}
  \textbf{\bibinfo{volume}{54}}, \bibinfo{pages}{1853 } (\bibinfo{year}{1996}).

\bibitem[{\citenamefont{Koulakov et~al.}(1996)\citenamefont{Koulakov, Fogler,
  and Shklovskii}}]{koulakov96}
\bibinfo{author}{\bibfnamefont{A.~A.} \bibnamefont{Koulakov}},
  \bibinfo{author}{\bibfnamefont{M.~M.} \bibnamefont{Fogler}},
  \bibnamefont{and} \bibinfo{author}{\bibfnamefont{B.~I.}
  \bibnamefont{Shklovskii}}, \bibinfo{journal}{Phys. Rev. Lett.}
  \textbf{\bibinfo{volume}{76}}, \bibinfo{pages}{499 } (\bibinfo{year}{1996}).

\bibitem[{\citenamefont{Lilly et~al.}(1999{\natexlab{a}})\citenamefont{Lilly,
  Cooper, Eisenstein, Pfeiffer, and West}}]{lilly99}
\bibinfo{author}{\bibfnamefont{M.~P.} \bibnamefont{Lilly}},
  \bibinfo{author}{\bibfnamefont{K.~B.} \bibnamefont{Cooper}},
  \bibinfo{author}{\bibfnamefont{J.~P.} \bibnamefont{Eisenstein}},
  \bibinfo{author}{\bibfnamefont{L.~N.} \bibnamefont{Pfeiffer}},
  \bibnamefont{and} \bibinfo{author}{\bibfnamefont{K.~W.} \bibnamefont{West}},
  \bibinfo{journal}{Phys. Rev. Lett.} \textbf{\bibinfo{volume}{82}}
  (\bibinfo{year}{1999}{\natexlab{a}}).

\bibitem[{\citenamefont{Du et~al.}(1999)\citenamefont{Du, Tsui, Stormer,
  Pfeiffer, Baldwin, and West}}]{du99}
\bibinfo{author}{\bibfnamefont{R.~R.} \bibnamefont{Du}},
  \bibinfo{author}{\bibfnamefont{D.~C.} \bibnamefont{Tsui}},
  \bibinfo{author}{\bibfnamefont{H.~L.} \bibnamefont{Stormer}},
  \bibinfo{author}{\bibfnamefont{L.~N.} \bibnamefont{Pfeiffer}},
  \bibinfo{author}{\bibfnamefont{K.~W.} \bibnamefont{Baldwin}},
  \bibnamefont{and} \bibinfo{author}{\bibfnamefont{K.~W.} \bibnamefont{West}},
  \bibinfo{journal}{Solid State Commun.} \textbf{\bibinfo{volume}{109}},
  \bibinfo{pages}{389 } (\bibinfo{year}{1999}).

\bibitem[{\citenamefont{Shayegan et~al.}(2000)\citenamefont{Shayegan,
  Manoharan, Papadakis, and Poortere}}]{shayegan00}
\bibinfo{author}{\bibfnamefont{M.}~\bibnamefont{Shayegan}},
  \bibinfo{author}{\bibfnamefont{H.~C.} \bibnamefont{Manoharan}},
  \bibinfo{author}{\bibfnamefont{S.~J.} \bibnamefont{Papadakis}},
  \bibnamefont{and} \bibinfo{author}{\bibfnamefont{E.~P.~D.}
  \bibnamefont{Poortere}}, \bibinfo{journal}{Physica E}
  \textbf{\bibinfo{volume}{6}}, \bibinfo{pages}{40 } (\bibinfo{year}{2000}).

\bibitem[{\citenamefont{Manfra et~al.}(2007)\citenamefont{Manfra, de~Picciotto,
  Jiang, Simon, Pfeiffer, West, and Sergent}}]{manfra07}
\bibinfo{author}{\bibfnamefont{M.~J.} \bibnamefont{Manfra}},
  \bibinfo{author}{\bibfnamefont{R.}~\bibnamefont{de~Picciotto}},
  \bibinfo{author}{\bibfnamefont{Z.}~\bibnamefont{Jiang}},
  \bibinfo{author}{\bibfnamefont{S.~H.} \bibnamefont{Simon}},
  \bibinfo{author}{\bibfnamefont{L.~N.} \bibnamefont{Pfeiffer}},
  \bibinfo{author}{\bibfnamefont{K.~W.} \bibnamefont{West}}, \bibnamefont{and}
  \bibinfo{author}{\bibfnamefont{A.~M.} \bibnamefont{Sergent}},
  \bibinfo{journal}{Phys. Rev. Lett.} \textbf{\bibinfo{volume}{98}},
  \bibinfo{pages}{206804} (\bibinfo{year}{2007}).

\bibitem[{\citenamefont{Kroemer}(1999)}]{kroemer99}
\bibinfo{author}{\bibfnamefont{H.}~\bibnamefont{Kroemer}},
  \bibinfo{journal}{cond-mat/9901016}  (\bibinfo{year}{1999}).

\bibitem[{\citenamefont{Takhtamirov and Volkov}(2000)}]{volkov00}
\bibinfo{author}{\bibfnamefont{E.~E.} \bibnamefont{Takhtamirov}}
  \bibnamefont{and} \bibinfo{author}{\bibfnamefont{V.~A.}
  \bibnamefont{Volkov}}, \bibinfo{journal}{JETP Lett.}
  \textbf{\bibinfo{volume}{71}}, \bibinfo{pages}{422} (\bibinfo{year}{2000}).

\bibitem[{\citenamefont{Rosenow and Stefan}(2001)}]{rosenow01}
\bibinfo{author}{\bibfnamefont{B.}~\bibnamefont{Rosenow}} \bibnamefont{and}
  \bibinfo{author}{\bibfnamefont{S.}~\bibnamefont{Stefan}},
  \bibinfo{journal}{International Journal of Modern Physics B}
  \textbf{\bibinfo{volume}{15}}, \bibinfo{pages}{1905} (\bibinfo{year}{2001}).

\bibitem[{\citenamefont{Fogler}(2002)}]{fogler-review}
\bibinfo{author}{\bibfnamefont{M.~M.} \bibnamefont{Fogler}},
  \emph{\bibinfo{title}{High Magnetic Fields: Applications in Condensed Matter
  Physics and Spectroscopy}}, vol. \bibinfo{volume}{595} of
  \emph{\bibinfo{series}{Lect. Notes Phys.}} (\bibinfo{publisher}{Springer},
  \bibinfo{address}{Berlin}, \bibinfo{year}{2002}).

\bibitem[{\citenamefont{Cooper et~al.}(2001)\citenamefont{Cooper, Lilly,
  Eisenstein, Jungwirth, Pfeiffer, and West}}]{cooper01}
\bibinfo{author}{\bibfnamefont{K.~B.} \bibnamefont{Cooper}},
  \bibinfo{author}{\bibfnamefont{M.~P.} \bibnamefont{Lilly}},
  \bibinfo{author}{\bibfnamefont{J.~P.} \bibnamefont{Eisenstein}},
  \bibinfo{author}{\bibfnamefont{T.}~\bibnamefont{Jungwirth}},
  \bibinfo{author}{\bibfnamefont{L.~N.} \bibnamefont{Pfeiffer}},
  \bibnamefont{and} \bibinfo{author}{\bibfnamefont{K.~W.} \bibnamefont{West}},
  \bibinfo{journal}{Sol. Stat. Comm.} \textbf{\bibinfo{volume}{119}},
  \bibinfo{pages}{89 } (\bibinfo{year}{2001}).

\bibitem[{\citenamefont{Rashba and Sherman}(1987)}]{rashba87}
\bibinfo{author}{\bibfnamefont{E.~I.} \bibnamefont{Rashba}} \bibnamefont{and}
  \bibinfo{author}{\bibfnamefont{E.~Y.} \bibnamefont{Sherman}},
  \bibinfo{journal}{Sov. Phys. Semicond.} \textbf{\bibinfo{volume}{21}},
  \bibinfo{pages}{1185} (\bibinfo{year}{1987}).

\bibitem[{\citenamefont{Sherman}(1995)}]{sherman95}
\bibinfo{author}{\bibfnamefont{E.~Y.} \bibnamefont{Sherman}},
  \bibinfo{journal}{Phys. Rev. B} \textbf{\bibinfo{volume}{52}},
  \bibinfo{pages}{1512} (\bibinfo{year}{1995}).

\bibitem[{\citenamefont{Fil}(2000)}]{fil00}
\bibinfo{author}{\bibfnamefont{D.~V.} \bibnamefont{Fil}}, \bibinfo{journal}{Low
  Temp. Phys.} \textbf{\bibinfo{volume}{26}}, \bibinfo{pages}{581 }
  (\bibinfo{year}{2000}).

\bibitem[{\citenamefont{Pan et~al.}(1999)\citenamefont{Pan, Du, Stormer, Tsui,
  Pfeiffer, Baldwin, and West}}]{pan99}
\bibinfo{author}{\bibfnamefont{W.}~\bibnamefont{Pan}},
  \bibinfo{author}{\bibfnamefont{R.}~\bibnamefont{Du}},
  \bibinfo{author}{\bibfnamefont{H.}~\bibnamefont{Stormer}},
  \bibinfo{author}{\bibfnamefont{D.}~\bibnamefont{Tsui}},
  \bibinfo{author}{\bibfnamefont{L.}~\bibnamefont{Pfeiffer}},
  \bibinfo{author}{\bibfnamefont{K.}~\bibnamefont{Baldwin}}, \bibnamefont{and}
  \bibinfo{author}{\bibfnamefont{K.}~\bibnamefont{West}},
  \bibinfo{journal}{Phys. Rev. Lett.} \textbf{\bibinfo{volume}{83}},
  \bibinfo{pages}{820 } (\bibinfo{year}{1999}).

\bibitem[{\citenamefont{Lilly et~al.}(1999{\natexlab{b}})\citenamefont{Lilly,
  Cooper, Eisenstein, Pfeiffer, and West}}]{lilly99a}
\bibinfo{author}{\bibfnamefont{M.~P.} \bibnamefont{Lilly}},
  \bibinfo{author}{\bibfnamefont{K.~B.} \bibnamefont{Cooper}},
  \bibinfo{author}{\bibfnamefont{J.~P.} \bibnamefont{Eisenstein}},
  \bibinfo{author}{\bibfnamefont{L.~N.} \bibnamefont{Pfeiffer}},
  \bibnamefont{and} \bibinfo{author}{\bibfnamefont{K.~W.} \bibnamefont{West}},
  \bibinfo{journal}{Phys. Rev. Lett.} \textbf{\bibinfo{volume}{83}},
  \bibinfo{pages}{824 } (\bibinfo{year}{1999}{\natexlab{b}}).

\bibitem[{\citenamefont{Jungwirth et~al.}(1999)\citenamefont{Jungwirth,
  MacDonald, Smr\ifmmode~\check{c}\else \v{c}\fi{}ka, and
  Girvin}}]{jungwirth99a}
\bibinfo{author}{\bibfnamefont{T.}~\bibnamefont{Jungwirth}},
  \bibinfo{author}{\bibfnamefont{A.~H.} \bibnamefont{MacDonald}},
  \bibinfo{author}{\bibfnamefont{L.}~\bibnamefont{Smr\ifmmode~\check{c}\else
  \v{c}\fi{}ka}}, \bibnamefont{and} \bibinfo{author}{\bibfnamefont{S.~M.}
  \bibnamefont{Girvin}}, \bibinfo{journal}{Phys. Rev. B}
  \textbf{\bibinfo{volume}{60}}, \bibinfo{pages}{15574} (\bibinfo{year}{1999}).

\bibitem[{\citenamefont{Stanescu et~al.}(2000)\citenamefont{Stanescu, Martin,
  and Phillips}}]{stanescu00}
\bibinfo{author}{\bibfnamefont{T.~D.} \bibnamefont{Stanescu}},
  \bibinfo{author}{\bibfnamefont{I.}~\bibnamefont{Martin}}, \bibnamefont{and}
  \bibinfo{author}{\bibfnamefont{P.}~\bibnamefont{Phillips}},
  \bibinfo{journal}{Phys. Rev. Lett.} \textbf{\bibinfo{volume}{84}},
  \bibinfo{pages}{1288} (\bibinfo{year}{2000}).

\bibitem[{\citenamefont{Manfra et~al.}(2005)\citenamefont{Manfra, Pfeiffer,
  West, de~Picciotto, and Baldwin}}]{manfra05}
\bibinfo{author}{\bibfnamefont{M.~J.} \bibnamefont{Manfra}},
  \bibinfo{author}{\bibfnamefont{L.~N.} \bibnamefont{Pfeiffer}},
  \bibinfo{author}{\bibfnamefont{K.~W.} \bibnamefont{West}},
  \bibinfo{author}{\bibfnamefont{R.}~\bibnamefont{de~Picciotto}},
  \bibnamefont{and} \bibinfo{author}{\bibfnamefont{K.~W.}
  \bibnamefont{Baldwin}}, \bibinfo{journal}{Appl. Phys. Lett.}
  \textbf{\bibinfo{volume}{86}}, \bibinfo{eid}{162106} (\bibinfo{year}{2005}).

\bibitem[{\citenamefont{Fung et~al.}(1997)\citenamefont{Fung, Cong, Albrecht,
  Nathan, Ruden, and Shtrikman}}]{fung97}
\bibinfo{author}{\bibfnamefont{A.~K.} \bibnamefont{Fung}},
  \bibinfo{author}{\bibfnamefont{L.}~\bibnamefont{Cong}},
  \bibinfo{author}{\bibfnamefont{J.~D.} \bibnamefont{Albrecht}},
  \bibinfo{author}{\bibfnamefont{M.~I.} \bibnamefont{Nathan}},
  \bibinfo{author}{\bibfnamefont{P.~P.} \bibnamefont{Ruden}}, \bibnamefont{and}
  \bibinfo{author}{\bibfnamefont{H.}~\bibnamefont{Shtrikman}},
  \bibinfo{journal}{J Appl. Phys.} \textbf{\bibinfo{volume}{81}},
  \bibinfo{pages}{502} (\bibinfo{year}{1997}).

\bibitem[{\citenamefont{Aleiner and Glazman}(1995)}]{aleiner95}
\bibinfo{author}{\bibfnamefont{I.~L.} \bibnamefont{Aleiner}} \bibnamefont{and}
  \bibinfo{author}{\bibfnamefont{L.~I.} \bibnamefont{Glazman}},
  \bibinfo{journal}{Phys. Rev. B} \textbf{\bibinfo{volume}{52}},
  \bibinfo{pages}{11296} (\bibinfo{year}{1995}).

\bibitem[{\citenamefont{Bir and Pikus}(1974)}]{birpikus74}
\bibinfo{author}{\bibfnamefont{G.~L.} \bibnamefont{Bir}} \bibnamefont{and}
  \bibinfo{author}{\bibfnamefont{G.~E.} \bibnamefont{Pikus}},
  \emph{\bibinfo{title}{Symmetry and strain-induced effects in semiconductors}}
  (\bibinfo{publisher}{Wiley}, \bibinfo{address}{New York},
  \bibinfo{year}{1974}).

\bibitem[{\citenamefont{Luttinger}(1956)}]{luttinger56}
\bibinfo{author}{\bibfnamefont{J.~M.} \bibnamefont{Luttinger}},
  \bibinfo{journal}{Phys. Rev.} \textbf{\bibinfo{volume}{102}},
  \bibinfo{pages}{1030} (\bibinfo{year}{1956}).

\bibitem[{\citenamefont{Nedorezov}(1971)}]{nedorezov71}
\bibinfo{author}{\bibfnamefont{S.~S.} \bibnamefont{Nedorezov}},
  \bibinfo{journal}{Soviet Physics--Solid State} \textbf{\bibinfo{volume}{12}},
  \bibinfo{pages}{1814} (\bibinfo{year}{1971}).

\bibitem[{\citenamefont{Kane}(1957)}]{kane57}
\bibinfo{author}{\bibfnamefont{E.~O.} \bibnamefont{Kane}},
  \bibinfo{journal}{Journal of Physics and Chemistry of Solids}
  \textbf{\bibinfo{volume}{1}}, \bibinfo{pages}{249} (\bibinfo{year}{1957}).

\bibitem[{\citenamefont{Kukushkin et~al.}(1988)\citenamefont{Kukushkin,
  Meshkov, and Timofeyev}}]{kukushkin88}
\bibinfo{author}{\bibfnamefont{I.~V.} \bibnamefont{Kukushkin}},
  \bibinfo{author}{\bibfnamefont{S.~V.} \bibnamefont{Meshkov}},
  \bibnamefont{and} \bibinfo{author}{\bibfnamefont{V.~B.}
  \bibnamefont{Timofeyev}}, \bibinfo{journal}{Uspehi Fiz. Nauk}
  \textbf{\bibinfo{volume}{155}}, \bibinfo{pages}{219} (\bibinfo{year}{1988}).

\bibitem[{\citenamefont{Zhu et~al.}(2002)\citenamefont{Zhu, Pan, Stormer,
  Pfeiffer, and West}}]{zhu02}
\bibinfo{author}{\bibfnamefont{J.}~\bibnamefont{Zhu}},
  \bibinfo{author}{\bibfnamefont{W.}~\bibnamefont{Pan}},
  \bibinfo{author}{\bibfnamefont{H.~L.} \bibnamefont{Stormer}},
  \bibinfo{author}{\bibfnamefont{L.~N.} \bibnamefont{Pfeiffer}},
  \bibnamefont{and} \bibinfo{author}{\bibfnamefont{K.~W.} \bibnamefont{West}},
  \bibinfo{journal}{Phys. Rev. Lett.} \textbf{\bibinfo{volume}{88}},
  \bibinfo{pages}{116803} (\bibinfo{year}{2002}).

\end{thebibliography}
\end{document}